\documentclass[preprint,12pt]{elsarticle}

\usepackage{graphicx}
\usepackage{amsthm,amsmath}
\usepackage{amssymb}
\usepackage{mathtools}
\newtheorem{Theorem}{Theorem}[section]
\newtheorem{Lemma}[Theorem]{Lemma}

\newtheorem{Proposition}[Theorem]{Proposition}

\newtheorem{Remark}{Remark}

\usepackage{mathtools}
\usepackage[toc,page]{appendix}
\usepackage[colorlinks]{hyperref}
\usepackage{bm}
\usepackage{graphicx}
\usepackage{amssymb}
\usepackage{lineno}

\journal{Journal Name}

\usepackage{lineno}
\usepackage{float}
\journal{Journal Name}

\begin{document}
\begin{frontmatter}

\title{A practical test for a planted community in heterogeneous networks}

\author[1]{Mingao Yuan\corref{cor1}}
\ead{mingao.yuan@ndsu.edu}

\author[1]{ Qian Wen}
\ead{qian.wen@ndsu.edu }

\cortext[cor1]{Corresponding author}
%\address[2]{Department of}
\address[1]{Department of Statistics, North Dakota State University, Fargo, ND,USA, 58102.}

\begin{abstract}
 One of the fundamental task in graph data mining is to find a planted community(dense subgraph), which has wide application in biology, finance, spam detection and so on. For a real network data, the existence of a dense subgraph is generally unknown. Statistical tests have been devised to testing the existence of dense subgraph in a homogeneous random graph. However, many networks present extreme heterogeneity, that is, the degrees of nodes or vertexes don't concentrate on a typical value. The existing tests designed for homogeneous random graph are not straightforwardly applicable to the heterogeneous case. Recently, scan test was proposed for detecting a dense subgraph in heterogeneous(inhomogeneous) graph(\cite{BCHV19}). However, the computational complexity of the scan test is generally not
polynomial in the graph size, which makes the test impractical for large or moderate networks. In this paper, we propose a polynomial-time test that has the standard normal distribution as the null limiting distribution. The power of the test is theoretically investigated and we evaluate the performance of the test by simulation and real data example.
\end{abstract}


\end{frontmatter}

\section{Introduction}

A graph(network) $\mathcal{G}=\left(V,E\right)$ consists of a node(vertex) set $V$ and an edge set $E$. It has been used to model many real world phenomena. For example, in a social network, an individual is a node and an edge between two nodes would represent the friendship. Usually, real network data presents community structure, that is, the nodes of the network can be clustered into groups, such that the nodes within a group is densely connected. A fundamental topic in analyzing the network data is to identify the densely connected subgraphs, which has wide applications in various fields, such as the spam detection in web graphs(\cite{GKT05}), detection of suspicious fake reviews in product review network(\cite{HSBSSF16}), and so on. In the community detection literature, various algorithms have been proposed to extract a dense subgraph in a network(\cite{AC09,C00,CS10,  GKT05, HWX2018,KS09, WLKN09}). For a given network, these algorithms typically output a dense subgraph, which is meaningful only if dense subgraph exists.  However, whether or not a dense subgraph exists in a given network is unknown in practice. This leads to the problem of detecting the presence of a dense subgraph(\cite{AV14,BCHV19,VA15}). This problem can be formulated as a statistical hypothesis testing: under the null hypothesis, there isn't any dense subgraph while under the alternative hypothesis, a dense subgraph is present. Now the question is how to devise a test statistic that can distinguish the alternative hypothesis from the null hypothesis with high power.

 Under the assumption that a given network is generated from a homogeneous random graph, in which the average degree of each node is the same, several powerful tests have been proposed in the literature, for example, the degree variance test and the scan test in (\cite{AV14,VA15}). In practice, many real networks display the inhomogeneity property, that is, the vertex degrees are significantly distinct. In many cases, the degrees would follow a power law(\cite{ACL01}). To accommodate the inhomogenity, the inhomogeneous random graph has been introduced to model the degree diversity(\cite{BJR07}). It's important to devise a test statistic to detect the presence of dense subgraph in inhomogeneous networks. In this case, however, it's a harder question due to the inhomogeneity. The test statistics designed for homogeneous networks can not be straightforwardly extended to the inhomogeneous case. To our knowledge, the only existing test statistic for a planted community is the  scan test proposed by(\cite{BCHV19}). However, there are two drawbacks that make the scan test impractical: the size of the planted community is assumed to be known and the computational complexity is not polynomial in the graph size.

In this paper, we propose a polynomial-time test statistic to detect the presence of a dense subgraph in inhomogeneous random graph. This test doesn't require the knowledge of the size of dense subgraph. Besides, the limiting distribution of the test is the standard normal distribution under the null hypothesis and the power of the test may approach one under the alternative hypothesis. We evaluate the performance of the test by simulation and real data application.

\section{The Model and Main result}
In this section, we introduce the model and present the main result. The inhomogeneous random graph
$\mathcal{G}\left(n, p, W\right)$ is defined as follows: given a nonnegative weight vector $W=\{W_{i}\}_{1\leq i\leq n}$, every pair of nodes $i,j$ in $\mathcal{G}\left(n, p, W\right)$ are joined as an edge with probability $pW_{i}W_{j}$ independently. The adjacency matrix $A$ of a graph is a symmetric $\left(0,1\right)$-matrix with zeros on its diagonal and $A_{ij}=1$ if $\left(i,j\right)$ is an edge, $A_{ij}=0$ otherwise. For $\mathcal{G}(n, p, W)$, the adjacency matrix $A$ is a symmetric random matrix, with elements following the independent Bernoulli distributions given $W$, that is,
\[\mathbb{P}\left(A_{ij}=1|W\right)=pW_{i}W_{j},\ \ 1\leq i<j\leq n,\] 
and $A_{ij}$ is independent of $A_{kl}$ if $\{i,j\}\neq\{k,l\}$.
If $W_{i}\equiv1\left(1\leq i\leq n\right)$, the random graph is the homogeneous Erd\"{o}s-R\'{e}nyi model, where nodes $i,j$ form an edge randomly with probability $p$(\cite{ER60}). 

The \textit{inhomogeneous dense subgraph model} $\mathcal{G}\left(n, r, W, \frac{a}{n},\frac{b}{n}\right)\left(a\geq b\right)$ is defined as follows. Let $Z=\left(Z_1,Z_2,\dots, Z_n\right)$ be a vector of independently and identically distributed Bernoulli random variables, with $\mathbb{P}\left(Z_i=1\right)=r$ $\left(i\in\{1,2,\dots,n\}\right)$. Then the edge probabilities are
\begin{equation}\label{network}
 \mathbb{P}\left(A_{ij}=1|Z,W\right)=W_iW_j\frac{\left(a-b\right)Z_iZ_j+b}{n}.
 \end{equation}
In this model, $Z_i=1$ indicates the node $i$ is selected to be a node in the dense subgraph. Two nodes randomly form an edge with probability $\frac{a}{n}W_iW_j$ if they are both in the dense subgraph, with probability $\frac{b}{n}W_iW_j$ otherwise. The parameter $r$ controls the size of the dense subgraph which contains $nr$ nodes on average. This model is called the rank-1 model in (\cite{BCHV19}). Besides, the homogeneous \textit{planted dense subgraph model} studied in (\cite{AV14,HWX2018,VA15}) is a special case of our model with $W_i\equiv1, i=1,2,\dots,n$. In this paper, the parameters $a,b$ and $p$ depend on $n$.

Whether a dense subgraph exists in a given network is statistically formulated as the following hypothesis testing problem
 \begin{equation}\label{H0H1}
     H_0: a=b,\hskip 1cm H_1: a>b. 
 \end{equation}
Under the null hypothesis $H_0$, the edge probabilities in (\ref{network}) are independent of $Z$, hence there isn't any dense subgraph in the network. Under the alternative hypothesis $H_1$, a dense subgraph is present. 

Due to the inhomogeneity, the testing problem (\ref{H0H1}) is much more difficult than the homogeneous case, since there are so many unknown parameters $a$, $b$ and $W_i$ $\left(i\in\{1, 2, \dots, n\}\right)$ in the model. To our knowledge, the only existing test is the scan test proposed in (\cite{BCHV19}), which assumes the model parameters are known and can't be computed in polynomial time. As a result, the scan test is not feasible for large or moderate networks. Therefore, it's necessary to devise a test statistic that has polynomial-time computational complexity and is free of any unknown model parameters. Our idea in this paper is to use the cycle-density statistic.

A $k$-cycle in a graph is a non-empty trail consisting of $k$-edges in which the only repeated vertices are the first and last vertices.
Define the density of 3-cycle as
\[\widehat{C}_3=\frac{1}{\binom{n}{3}}\sum_{1\leq i<j<k\leq n}A_{ij}A_{jk}A_{ki},\]
Let $S$ be the set of all the permutations on index set $\{i_1,i_2,\dots,i_6\}$ for $i_s\neq i_t(1\leq s<t\leq 6)$. Define the 6-cycle density as
\begin{eqnarray*}
\widehat{C}_6
&=&\frac{1}{6!\binom{n}{6}}\sum_{1\leq i_1<
\dots<i_6\leq n}\sum_{\pi\in S}A_{\pi_{i_1}\pi_{i_2}}A_{\pi_{i_2}\pi_{i_3}}A_{\pi_{i_3}\pi_{i_4}}\\
&&\times A_{\pi_{i_4}\pi_{i_5}}A_{\pi_{i_5}\pi_{i_6}}A_{\pi_{i_6}\pi_{i_1}}.
\end{eqnarray*}

We propose the cycle-count test statistic as
\begin{equation}\label{tstat}
    T_n=\frac{\sqrt{\binom{n}{3}}\big(\widehat{C}_3^2-\widehat{C}_6\big)}{2\widehat{C}_3\sqrt{\widehat{C}_3}}.
\end{equation}
Note that the computational complexity of $T_n$ is at most $O(n^6)$, which is polynomial in graph size and can be significantly reduced if the network is relatively sparse.

 \begin{Remark}
 Similar idea was used in \cite{AV14} to propose
 the degree variance test, which is a function of the number of 2-path instead of cycle. The degree variance test is not applicable in the inhomogeneous case \cite{BCHV19}. 
 \end{Remark}

 \subsection{Rationale of the Cycle-Count Test Statistic}
  For convenience, denote $\alpha_n\asymp \beta_n$ if $0<c_1\leq\alpha_n/\beta_n\leq c_2<\infty$ for some constants $c_1$ and $c_2$; $\alpha_n\ll\beta_n$ if $\alpha_n=o(\beta_n)$. 
  
 Let $C_3=\mathbb{E}\widehat{C}_3$ and $C_6=\mathbb{E}\widehat{C}_6$. Define $\Lambda_1=C_3^2-C_6$ under $H_1$ and $\Lambda_0=C_3^2-C_6$ under $H_0$. Let $||W||_2$ be the Euclidean norm of the vector $W$. Then under $H_1$ and some mild conditions on $W$, it yields
 \begin{equation}\label{c3w}
    C_3=\frac{A_n}{3!\binom{n}{3}}\|W\|_2^6\left(1+o\left(1\right)\right),
\end{equation}
\begin{eqnarray}\nonumber
C_6&=&\frac{B_n}{6!\binom{n}{6}}\|W\|_2^{12}\left(1+o\left(1\right)\right),
\end{eqnarray}
where $A_n$ and $B_n$ depend on $a,b,r,n$ and are given in the proof of Proposition \ref{diff}.
 
 \begin{Proposition}\label{diff}
  Under $H_1$, it follows that
   \[\Lambda_1=\frac{2\left(a-b\right)^3b^3r^3\left(1-r\right)^3}{n^6}\frac{\|W\|_2^{12}}{3!^2\binom{n}{3}^2}(1+o(1))+\frac{B_n\|W\|_2^{12}}{n^6}O\Big(\frac{1}{n}\Big).\]
   Under $H_0$,  $\Lambda_0=\frac{B_n\|W\|_2^{12}}{n^6}O\Big(\frac{1}{n}\\Big)$, since $a=b$. Hence if $a\asymp b$ and $1\ll nr^3$, then $\Lambda_0=o\left(\Lambda_1\right)$.
 \end{Proposition}

Based on Proposition \ref{diff}, $\Lambda_1$ is of higher order than $\Lambda_0$ and hence the quantity $C_3^2-C_6$ characterizes the difference between $H_1$ and $H_0$. The proposed test statistic $T_n$ in (\ref{tstat}) exploited this fact. The numerator of $T_n$ is just an empirical estimator of this quantity. To derive the asymptotic distribution, we scale it as in (\ref{tstat}).

 \subsection{Main Results}
 Let $p_0=\frac{a}{n}$ and $||W||_k^k=\sum_{i=1}^nW_i^k$ for any positive integer $k$.
The following theorem gives the limiting distribution of $T_n$ under $H_0$. 

\begin{Theorem}\label{tm1}
Suppose $1\ll\|W\|_k^k\leq c_1\|W\|_2^2\leq c_2n$ for integer $k$ between 3 and 12 and some constants $c_1,c_2$. If $1\ll np_0\ll \sqrt{n}$ and $1\ll p_0\|W\|_2^2$, then
under the null hypothesis $H_0$, $T_n$ converges in law to the standard normal distribution.
\end{Theorem}

Based on Theorem \ref{tm1}, it's easy to calibrate the test statistic since the limiting distribution doesn't contain any unknown parameters. For a given significance level $\alpha$, reject $H_0$ if $|T_n|>Z_{\frac{\alpha}{2}}$, where $Z_{\frac{\alpha}{2}}$ is the $(1-\frac{\alpha}{2})100\%$ quantile of the standard normal distribution.

\begin{Theorem}\label{tm2}
Under the alternative hypothesis $H_1$ and the conditions of Theorem \ref{tm1},
 the test statistic $T_n=\lambda_n+O_p(1)$, where
\begin{equation}\label{powerc}
\lambda_n^2=\frac{(a-b)^6b^6r^6(1-r)^6\|W\|_2^6}{n^{12}p_0^9}.
\end{equation}
Hence the power approaches one if $\lambda_n$ goes to infinity. If $a\asymp b$,  $\lambda_n\rightarrow\infty$ is equivalent to
\begin{equation}\label{credced}
\lim_{n\rightarrow\infty}\frac{r^2(a-b)\|W\|_2^2}{n}=\infty.
\end{equation}
\end{Theorem}

According to (\ref{credced}), the degree diversity parameter $\|W\|_2$, size of the dense subgraph(on average, there are $nr$ nodes in the dense subgraph)  and the difference between $a$ and $b$ jointly control the power of the test.

\begin{Remark}
The power of the proposed test may still go to one even when the size of a dense subgraph is not linear in the graph size. To see this, suppose there are $n^{\epsilon}(0<\epsilon<1)$ nodes in the dense subgraph on average. 
Let $\|W\|_2^2=n$ and  $a-b=n^{\eta}$, for $\eta\in (0, 0.5)$. Then if $r\gg n^{-\frac{\eta}{2}}$, the power goes to one by (\ref{credced}). In this case, the dense subgraph consists of $nr\in(n^{1-\frac{\eta}{2}},n)$ nodes on average.
\end{Remark}

\subsection{An example}
In this subsection, we give one example that satisfy the conditions of Theorem \ref{tm1}.
%\subsection{Example 1}
Let $W_i=\frac{i}{n}$, $i\in\{1,2,\dots,n\}$. Then straightforward computation yields $||W||_2^2=\frac{n}{3}(1+O(1))$ and $||W||_k^k=\frac{n}{k+1}(1+O(1))$ for integer $k\in\{2,3,\dots,12\}$. If $1\ll np_0\ll \sqrt{n}$, all conditions in Theorem \ref{tm1} are satisfied. Note that this model is highly inhomogeneous. The average degree $d_i$ of vertex $i$ is given by
\[d_i=\sum_{j}p_0W_iW_j=\frac{ip_0}{2}+\frac{ip_0}{2n}-\frac{i^2p_0}{n^2},\hskip 5mm i=1,2,\dots,n,\]
where $p_0=\frac{a}{n}=\frac{b}{n}$ under $H_0$.
The expected degree $d_n$ of node $n$ is aproximately $n$ times the expected degree $d_1$ of node 1, that is, $d_n= nd_1(1+o(1))$.  Hence this model can produce graphs with highly heterogeneous degree sequence and is essentially different from the homogeneous Erd\"{o}s-R\'{e}nyi model, where all the average degrees are the same.

%\subsection{Example 2}

%Let $W_i$ $(i\in\{1,2,\dots,n\})$ be generated from the uniform distribution defined on the interval $[1,c]$, that is, $W_i\sim Uniform(1,c)$ and fix it as weight. Then $n\leq||W||_k^k\leq c^kn$. If $1\ll np_0\ll \sqrt{n}$, Theorem \ref{tm1} applies to this model, which is inhomogeneous if $c>1$. 

\section{Experiments}

We run a small simulation and use real data to evaluate the performance of the proposed test.

\subsection{Simulation}
Fix $n=400$ and the significance level $\alpha=0.05$ throughout this simulation. We repeat the experiment 200 times to calculate the empirical size and power of the proposed test $T_n$.

%Firstly,

We run simulation under the setting of the example given in section 2.3 , that is, $W_i=\frac{i}{n}$, $i\in\{1,2,\dots,n\}$. Under $H_0$, take $p_0=0.08$. Under $H_1$, the parameters $a$, $b$ and $r$ are specified in Table \ref{pow2}. Generate the random vector $Z$ from the Bernoulli distribution with parameter $r$, i.e, $Z_i\sim Bernoulli(r)$, $i\in\{1,2,\dots,n\}$. Then the adjacency matrix $A$ is generated by (\ref{network}). The empirical size and powers are summarized in Table \ref{pow2}. The size is 0.03, less than the nominal level 0.05. As the difference $\frac{a-b}{n}$ enlarges, the power increases for each fixed $r$. For fixed $\frac{a}{n}$ and $\frac{n}{n}$, as $r$ increases, the power gets larger. The largest powers are greater than 0.9. All these findings are consistent with Theorem \ref{tm1} and Theorem \ref{tm2}.

\begin{table}[ht]
\caption{Size and powers of $T_n$.}
\centering
\begin{tabular}{|p{2.5cm} |p{1.5cm}|p{1.5cm}|p{1.5cm}| p{1.5cm}| }
\hline
$(\frac{a}{n},\frac{b}{n})$&$r=0.1$ & $r=0.2$ &$r=0.3$\\ 
\hline
\hline
(0.08, 0.08)& 0.03 &0.03   & 0.03        \\
(0.48, 0.08)& 0.19   & 0.63   & 0.67          \\
(0.56, 0.08)& 0.33   &  0.93   &  0.95         \\
\hline
\end{tabular}\label{pow2}
\end{table}

 \subsection{Application to Real Data}
 Note that the proposed test statistic $T_n$ doesn't depend on any unknown model parameters and can be computed in polynomial-time,  it's convenient to apply it to real networks. Here, we consider the three real networks: political book, dolphins, and road-chesapeake. They are publicly available at \url{http://networkrepository.com}. The basic summary statistics and the calculated test statistic $T_n$ in (\ref{tstat}) are listed in Table \ref{power2}. 

\begin{table}[h]
\caption{Data sets and the calucalted test statistic $T_n$}
\centering
\begin{tabular}{|p{3cm} |p{1.8cm}|p{1.8cm}|p{1.8cm}|}
\hline
Dataset & \# Nodes & \# Edges & $T_n$ \\ 
\hline
\hline
 political book& 105 &441   &  9.82 \\
dolphins& 62    &  159    &  4.19\\
road-chesapeake& 39 &170   &  6.53  \\
\hline
\end{tabular}\label{power2}
\end{table}
All the calculated test statistics $T_n$ are greater than 1.96, we reject the null hypothesis $H_0$ and conclude that there exists a dense subgraph in each of these networks at significance level 0.05.

\section{Proof of  Main Results}
In this section, we prove Theorem \ref{tm1} and Theorem \ref{tm2}. For convenience, denote $\alpha_n\asymp \beta_n$ if $0<c_1\leq\alpha_n/\beta_n\leq c_2<\infty$ for some constants $c_1$ and $c_2$; $\alpha_n\ll\beta_n$ if $\alpha_n=o(\beta_n)$.

\subsection{Proof of Proposition \ref{diff}}

Under $H_1$, direct computation yields
\begin{eqnarray*}
   &&\mathbb{E}(A_{12}A_{23}A_{31})\\
    &=&\mathbb{E}\big[\mathbb{P}(A_{12}A_{23}A_{31}=1|Z)\big]\\
    &=&\mathbb{E}\Big[W_1^2W_2^2W_3^2\frac{(a-b)Z_1Z_2+b}{n}\frac{(a-b)Z_2Z_3+b}{n}\frac{(a-b)Z_3Z_1+b}{n}\Big]\\
        &=&A_nW_1^2W_2^2W_3^2,
\end{eqnarray*}
and
\begin{eqnarray*}
  \mathbb{E}(A_{12}A_{23}A_{34}A_{45}A_{56}A_{61})
    &=&\mathbb{E}\big[\mathbb{P}(A_{12}A_{23}A_{34}A_{45}A_{56}A_{61}=1|Z)\big]\\
    &=&B_nW_1^2W_2^2W_3^2W_4^2W_5^2W_6^2,
\end{eqnarray*}

where 
\begin{eqnarray*}
A_n&=&\frac{(a-b)^3r^3+3(a-b)^2br^3+3(a-b)b^2r^2+b^3}{n^3},\\
B_n&=&\Big(\frac{r^6\big[(a-b)^6+6(a-b)^5b+9(a-b)^4b^2+2(a-b)^3b^3\big]}{n^6}\\
    &&+\frac{r^5\big[6(a-b)^4b^2+12(a-b)^3b^3\big]}{n^6}\\
    &&+\frac{r^4\big[6(a-b)^3b^3+9(a-b)^2b^4\big]}{n^6}\\
    &&+\frac{r^36(a-b)^2b^4}{n^6}+\frac{r^26(a-b)b^5}{n^6}+\frac{b^6}{n^6}\Big).
\end{eqnarray*}

Hence,
\begin{eqnarray}\nonumber
C_3&=&\mathbb{E}\widehat{C}_3=
\frac{1}{\binom{n}{3}}\sum_{1\leq i<j<k\leq n}\mathbb{E}A_{ij}A_{jk}A_{ki}\\   \nonumber
&=&\frac{1}{\binom{n}{3}}\sum_{1\leq i<j<k\leq n}W_i^2W_j^2W_k^2\mathbb{E}\Big[\frac{(a-b)Z_iZ_j+b}{n}\frac{(a-b)Z_jZ_k+b}{n}\\ \nonumber
&&\times\frac{(a-b)Z_kZ_i+b}{n}\Big]\\   \nonumber
&=&\frac{A_n}{\binom{n}{3}}\sum_{1\leq i<j<k\leq n}W_i^2W_j^2W_k^2.\\   \nonumber
\end{eqnarray}

For generic constants $c_1$ and $c_2$, it follows
\begin{eqnarray*}  \nonumber
\sum_{1\leq i,j,k\leq n}W_i^2W_j^2W_k^2&=&3!\sum_{1\leq i<j<k\leq n}W_i^2W_j^2W_k^2+c_1\sum_{1\leq i,j\leq n}W_i^4W_j^2\\ 
&&+c_2\sum_{1\leq i\leq n}W_i^6.
\end{eqnarray*}
Hence, 
\begin{eqnarray} \nonumber
\sum_{1\leq i<j<k\leq n}W_i^2W_j^2W_k^2&=&\frac{1}{3!}\|W\|_2^6-c_1\|W\|_4^4\|W\|_2^2-c_2\|W\|_6^6\\
&=&\frac{1}{3!}\|W\|_2^6(1+o(1)),
\end{eqnarray}
if $\|W\|_k^k\leq C\|W\|_2^2$ for  integer $k$ between 4 and 12 and some constant $C$, and $\|W\|_2\rightarrow\infty$. Then
\begin{equation}\label{c3w}
    C_3=\frac{A_n}{3!\binom{n}{3}}\|W\|_2^6(1+o(1)).
\end{equation}

A similar calculation yields
\begin{eqnarray}\nonumber
C_6&=&\mathbb{E}\widehat{C}_6=
\frac{1}{\binom{n}{6}}\sum_{1\leq i_1<i_2<
\dots<i_6\leq n}\mathbb{E}A_{i_1i_2}A_{i_2i_3}A_{i_3i_4}A_{i_4i_5}A_{i_5i_6}A_{i_6i_1}\\  \nonumber
&=&\frac{B_n}{\binom{n}{6}}\sum_{1\leq i_1<i_2<
\dots<i_6\leq n}W_{i_1}^2W_{i_2}^2W_{i_3}^2W_{i_4}^2W_{i_5}^2W_{i_6}^2.
\end{eqnarray}
It's easy to verify that
\begin{eqnarray*}
&&\sum_{1\leq i_1,i_2,
\dots,i_6\leq n}W_{i_1}^2W_{i_2}^2W_{i_3}^2W_{i_4}^2W_{i_5}^2W_{i_6}^2\\
&=&6!\sum_{1\leq i_1<i_2<
\dots<i_6\leq n}W_{i_1}^2W_{i_2}^2W_{i_3}^2W_{i_4}^2W_{i_5}^2W_{i_6}^2\\
&&+\sum_{\substack{1\leq i_1,i_2,
\dots,i_5\leq n\\ 0\leq r_1,r_2,\dots,r_5\leq 6}}c_{r_1r_2r_3r_4r_5}W_{i_1}^{2r_1}W_{i_2}^{2r_2}W_{i_3}^{2r_3}W_{i_4}^{2r_4}W_{i_5}^{2r_5},
\end{eqnarray*}
where $r_k(1\leq k\leq 6)$ are integers and $r_1+r_2+r_3+r_4+r_5=6$ in the second summation term and $c_{r_1r_2r_3r_4r_5}$ are constants dependent on $r_1,\dots,r_5$. Then
\begin{eqnarray}\nonumber
C_6&=&\frac{B_n}{6!\binom{n}{6}}\|W\|_2^{12}(1+o(1)),
\end{eqnarray}
if $\|W\|_k^k\leq C\|W\|_2^2$ for integer $k$ between 3 and 12 and some constant $C$, and $\|W\|_2\rightarrow\infty$.

Let $\Lambda_1=C_3^2-C_6$ under $H_1$. Then
\begin{eqnarray*}
\Lambda_1&=&\frac{A_n^2}{3!^2\binom{n}{3}^2}\|W\|_2^{12}(1+o(1))-\frac{B_n}{6!\binom{n}{6}}\|W\|_2^{12}(1+o(1))\\
&=&\frac{A_n^2-B_n}{3!^2\binom{n}{3}^2}\|W\|_2^{12}(1+o(1))\\
&&+B_n\Big(\frac{1}{3!^2\binom{n}{3}^2}-\frac{1}{6!\binom{n}{6}}\Big)\|W\|_2^{12}(1+o(1))\\
&=&\frac{2(a-b)^3b^3r^3(1-r)^3}{n^6}\frac{\|W\|_2^{12}}{3!^2\binom{n}{3}^2}(1+o(1))\\
&&+\frac{B_n\|W\|_2^{12}}{n^6}O\Big(\frac{1}{n}\Big).
\end{eqnarray*}
Let $T_0=C_3^2-C_6$ under $H_0: a=b$. 
Then $T_0=\frac{B_n\|W\|_2^{12}}{n^6}O\Big(\frac{1}{n}\Big)$. If $a\asymp b$ and $r\gg n^{-\frac{1}{3}}$, then $T_0=o(T)$. 

\qed

\subsection{Proof of Theorem \ref{tm1}}

Write
\begin{eqnarray}\label{c3sc61}
\widehat{C}_3^2-\widehat{C}_6=(\widehat{C}_3^2-C_3^2)+(C_3^2-C_6)+(C_6-\widehat{C}_6).
\end{eqnarray}

Note that under $H_0$,
\begin{eqnarray}\nonumber
&&\widehat{C}_3-C_3\\  \nonumber
&=&
\frac{1}{\binom{n}{3}}\sum_{1\leq i<j<k\leq n}\big(A_{ij}A_{jk}A_{ki}-W_i^2W_j^2W_k^2p_0^3\big)\\   \nonumber
&=&\frac{1}{\binom{n}{3}}\sum_{1\leq i<j<k\leq n}\big(A_{ij}-W_iW_jp_0\big)\big(A_{jk}-W_jW_kp_0\big)\big(A_{ki}-W_kW_ip_0\big)\\   \nonumber
&&+\frac{1}{\binom{n}{3}}\sum_{1\leq i<j<k\leq n}\big(A_{ij}-W_iW_jp_0\big)\big(A_{jk}-W_jW_kp_0\big)W_kW_ip_0\\   \nonumber
&&+\frac{1}{\binom{n}{3}}\sum_{1\leq i<j<k\leq n}\big(A_{ij}-W_iW_jp_0\big)\big(A_{ki}-W_kW_ip_0\big)W_jW_kp_0\\   \nonumber
&&+ \frac{1}{\binom{n}{3}}\sum_{1\leq i<j<k\leq n}\big(A_{jk}-W_jW_kp_0\big)\big(A_{ki}-W_kW_ip_0\big)W_iW_jp_0\\   \nonumber
&&+\frac{1}{\binom{n}{3}}\sum_{1\leq i<j<k\leq n}\big(A_{ij}-W_iW_jp_0\big)W_jW_kp_0W_kW_ip_0\\   \nonumber
&&+\frac{1}{\binom{n}{3}}\sum_{1\leq i<j<k\leq n}W_iW_jp_0\big(A_{jk}-W_jW_kp_0\big)W_kW_ip_0\\   \label{c31}
&&+\frac{1}{\binom{n}{3}}\sum_{1\leq i<j<k\leq n}W_iW_jp_0W_jW_kp_0\big(A_{ki}-W_kW_ip_0\big).
\end{eqnarray}

Note that under $H_0$, $A_{ij}$ and $A_{kl}$ are independent if $\{i,j\}\neq\{k,l\}$. Then 
\[\mathbb{E}\big[\big(A_{ij}-W_iW_jp_0\big)\big(A_{kl}-W_kW_lp_0\big)\big]=0.\]
Hence, it's easy to get
\begin{eqnarray}\nonumber
&& \mathbb{E}\Big[\frac{1}{\binom{n}{3}}\sum_{1\leq i<j<k\leq n}\big(A_{ij}-W_iW_jp_0\big)\big(A_{jk}-W_jW_kp_0\big)\big(A_{ki}-W_kW_ip_0\big)\Big]^2\\   \nonumber
&=&\frac{1}{\binom{n}{3}^2}\sum_{\substack{1\leq i<j<k\leq n\\1\leq i_1<j_1<k_1\leq n }}\mathbb{E}\Big[\big(A_{ij}-W_iW_jp_0\big)\big(A_{jk}-W_jW_kp_0\big)\big(A_{ki}-W_kW_ip_0\big)\\   \nonumber
&&\times \big(A_{i_1j_1}-W_{i_1}W_{j_1}p_0\big)\big(A_{j_1k_1}-W_{j_1}W_{k_1}p_0\big)\big(A_{k_1i_1}-W_{k_1}W_{i_1}p_0\big)\Big]\\    \nonumber
&=&\frac{1}{\binom{n}{3}^2}\sum_{\substack{1\leq i<j<k\leq n }}\mathbb{E}\Big[\big(A_{ij}-W_iW_jp_0\big)^2\big(A_{jk}-W_jW_kp_0\big)^2\big(A_{ki}-W_kW_ip_0\big)^2\Big]\\   \nonumber
&=&\frac{1}{\binom{n}{3}^2}\sum_{\substack{1\leq i<j<k\leq n }}W_iW_jp_0\big(1-W_iW_jp_0\big)W_jW_kp_0\big(1-W_jW_kp_0\big)\\ \nonumber
&&\times W_kW_ip_0\big(1-W_kW_ip_0\big)\\   \label{c32}
&\asymp&\frac{p_0^3}{\binom{n}{3}^2}\sum_{\substack{1\leq i<j<k\leq n }}W_i^2W_j^2W_k^2=\frac{p_0^3}{3!\binom{n}{3}^2}\|W\|_2^6(1+o(1)),
\end{eqnarray}
if $\|W\|_k^k\leq C\|W\|_2^2$ for integer $k$ between 3 and 12 and some constant $C$, and $\|W\|_2\rightarrow\infty$.

\begin{eqnarray}\nonumber
&&\mathbb{E}\Big[\frac{1}{\binom{n}{3}}\sum_{1\leq i<j<k\leq n}\big(A_{ij}-W_iW_jp_0\big)\big(A_{jk}-W_jW_kp_0\big)W_kW_ip_0\Big]^2\\   \nonumber
&=&\frac{1}{\binom{n}{3}^2}\sum_{\substack{1\leq i<j<k\leq n\\1\leq i_1<j_1<k_1\leq n }}\mathbb{E}\Big[\big(A_{ij}-W_iW_jp_0\big)\big(A_{jk}-W_jW_kp_0\big)W_kW_ip_0\\  \nonumber
&&\times \big(A_{i_1j_1}-W_{i_1}W_{j_1}p_0\big)\big(A_{j_1k_1}-W_{j_1}W_{k_1}p_0\big)W_{k_1}W_{i_1}p_0\Big]\\   \nonumber
&=&\frac{1}{\binom{n}{3}^2}\sum_{\substack{1\leq i<j<k\leq n }}\mathbb{E}\Big[\big(A_{ij}-W_iW_jp_0\big)^2\big(A_{jk}-W_jW_kp_0\big)^2\\  \nonumber
&&\times W_k^2W_i^2p_0^2\Big]\\   \nonumber
&=&\frac{1}{\binom{n}{3}^2}\sum_{\substack{1\leq i<j<k\leq n }}\Big[W_iW_jp_0\big(1-W_iW_jp_0\big)W_jW_kp_0\\   \nonumber
&&\times\big(1-W_jW_kp_0\big)W_k^2W_i^2p_0^2\Big]\\   \label{c33}
&\asymp&\frac{p_0^4}{\binom{n}{3}^2}\sum_{\substack{1\leq i<j<k\leq n }}W_i^3W_j^2W_k^3=\frac{p_0^4}{3!\binom{n}{3}^2}\|W\|_2^6O(1).
\end{eqnarray}
if $\|W\|_k^k\leq C\|W\|_2^2$ for integer $k$ between 3 and 12 and some constant $C$, and $\|W\|_2\rightarrow\infty$.

\begin{eqnarray}\nonumber
&&\mathbb{E}\Big[\frac{1}{\binom{n}{3}}\sum_{1\leq i<j<k\leq n}\big(A_{ij}-W_iW_jp_0\big)W_jW_kp_0W_kW_ip_0\Big]^2\\   \nonumber
&=&\frac{1}{\binom{n}{3}^2}\sum_{\substack{1\leq i<j<k\leq n\\1\leq i_1<j_1<k_1\leq n }}\mathbb{E}\Big[\big(A_{ij}-W_iW_jp_0\big)W_jW_kp_0W_kW_ip_0\\   \nonumber
&&\times \big(A_{i_1j_1}-W_{i_1}W_{j_1}p_0\big)W_{j_1}W_{k_1}p_0W_{k_1}W_{i_1}p_0\Big]\\   \nonumber
&=&\frac{1}{\binom{n}{3}^2}\sum_{\substack{1\leq i<j<k\leq n\\1\leq i<j<k_1\leq n }}\mathbb{E}\Big[\big(A_{ij}-W_iW_jp_0\big)^2W_jW_kp_0W_kW_ip_0W_{j}W_{k_1}\\  \nonumber
&&\times p_0W_{k_1}W_{i}p_0\Big]\\   \nonumber
&=&\frac{1}{\binom{n}{3}^2}\sum_{\substack{1\leq i<j<k\leq n\\1\leq i<j<k_1\leq n }}\Big[W_iW_jp_0\big(1-W_iW_jp_0\big)W_jW_kp_0W_kW_i\\  \nonumber
&&\times p_0W_{j}W_{k_1}p_0W_{k_1}W_{i}p_0\Big]\\    \label{c34}
&\asymp&\frac{p_0^5}{\binom{n}{3}^2}\sum_{\substack{1\leq i<j<k\leq n\\1\leq i<j<k_1\leq n }}W_i^3W_j^3W_k^2W_{k_1}^2=\frac{p_0^5}{\binom{n}{3}^2}\|W\|_2^{8}O(1),
\end{eqnarray}
if $\|W\|_k^k\leq C\|W\|_2^2$ for integer $k$ between 3 and 12 and some constant $C$, and $\|W\|_2\rightarrow\infty$.

Under condition of Theorem \ref{tm1}, $p_0\|W\|_2=o(1)$, by  (\ref{c31}), (\ref{c32}), (\ref{c33}), (\ref{c34}), it follows

\begin{eqnarray}\nonumber
\widehat{C}_3-C_3&=&\frac{1}{\binom{n}{3}}\sum_{1\leq i<j<k\leq n}\big(A_{ij}-W_iW_jp_0\big)\big(A_{jk}-W_jW_kp_0\big)\\  \nonumber
&&\times\big(A_{ki}-W_kW_ip_0\big)+o_p\Big(\sqrt{\frac{p_0^3}{n^6}}\|W\|_2^3\Big),\\ \label{hatcc3}
\end{eqnarray}
and
\begin{eqnarray}\label{hatc3s}
\mathbb{E}\big[\widehat{C}_3-C_3\big]^2=\frac{p_0^3}{3!\binom{n}{3}^2}\|W\|_2^6(1+o(1)).
\end{eqnarray}
Then 
\begin{eqnarray}\nonumber
\widehat{C}_3^2-C_3^2&=&2C_3\big(\widehat{C}_3-C_3\big)+\big(\widehat{C}_3-C_3\big)^2\\  \label{c3c3}
&=&O_p\Big(\frac{A_n||W||_2^6}{n^3}\sqrt{\frac{p_0^3}{n^6}\|W\|_2^6}\Big)+O_p\Big(\frac{p_0^3}{n^6}\|W\|_2^6\Big),
\end{eqnarray}
and the leading term is $C_3\big(\widehat{C}_3-C_3\big)$ if $p_0||W||_2^2\rightarrow\infty$.

For the identity permutation $\pi$ in $\widehat{C}_6-C_6$, we have
\begin{eqnarray}\nonumber
&&\widehat{C}_6-C_6\\  \nonumber
&=&\frac{1}{6!\binom{n}{6}}\sum_{1\leq i_1<i_2<
\dots<i_6\leq n}\big(A_{i_1i_2}A_{i_2i_3}A_{i_3i_4}A_{i_4i_5}A_{i_5i_6}A_{i_6i_1}\\  \nonumber
&&-W_{i_1}^2W_{i_2}^2W_{i_3}^2W_{i_4}^2W_{i_5}^2W_{i_6}^2p_0^6\big)\\   \label{c6}
\end{eqnarray}

If $\|W\|_k^k\leq C\|W\|_2^2$ for integer $k$ between 3 and 12 and some constant $C$, and $\|W\|_2\rightarrow\infty$, it's easy to verify that
\begin{eqnarray}\nonumber
&& \mathbb{E}\Big[\frac{1}{\binom{n}{6}}\sum_{1\leq i_1<\dots<i_6\leq n}\big(A_{i_1i_2}-W_{i_1}W_{i_2}p_0\big)\big(A_{i_2i_3}-W_{i_2}W_{i_3}p_0\big)\\ \nonumber
&&\times\big(A_{i_3i_4}-W_{i_3}W_{i_4}p_0\big)\\  \label{c6w1}
&&\times\big(A_{i_4i_5}-W_{i_4}W_{i_5}p_0\big)\big(A_{i_5i_6}-W_{i_5}W_{i_6}p_0\big)\big(A_{i_6i_1}-W_{i_6}W_{i_1}p_0\big) \Big]^2\\   \nonumber
&=&\frac{p_0^6}{\binom{n}{6}^2}\sum_{1\leq i_1<\dots<i_6\leq n}W_{i_1}^2W_{i_2}^2W_{i_3}^2W_{i_4}^2W_{i_5}^2W_{i_6}^2=\frac{p_0^6}{6!\binom{n}{6}^2}\|W\|_2^{12}(1+o(1)),
\end{eqnarray}

\begin{eqnarray}\nonumber
&&\mathbb{E}\Big[\frac{1}{\binom{n}{6}}\sum_{1\leq i_1<i_2<
\dots<i_6\leq n}\big(A_{i_1i_2}-W_{i_1}W_{i_2})W_{i_2}W_{i_3}^2W_{i_4}^2W_{i_5}^2W_{i_6}^2W_{i_1}p_0^5\big)\Big]^2\\   \nonumber
&=&\frac{1}{\binom{n}{6}^2}\sum_{\substack{1\leq i_1<i_2<
\dots<i_6\leq n\\ 1\leq j_1<j_2<
\dots<j_6\leq n}}\mathbb{E}\Big[\big(A_{i_1i_2}-W_{i_1}W_{i_2})W_{i_2}W_{i_3}^2W_{i_4}^2W_{i_5}^2W_{i_6}^2W_{i_1}p_0^5\big)\\   \nonumber
&&\times \big(A_{j_1j_2}-W_{j_1}W_{j_2})W_{j_2}W_{j_3}^2W_{j_4}^2W_{j_5}^2W_{j_6}^2W_{j_1}p_0^5\big)\Big]\\   \nonumber
&=&\frac{p_0^{11}}{\binom{n}{6}^2}\sum_{\substack{1\leq i_1<i_2<
\dots<i_6\leq n\\ 1\leq j_3<
\dots<j_6\leq n}}W_{i_1}^3W_{i_2}^3\big(1-W_{i_1}W_{i_2}p_0)\\ \nonumber
&&\times W_{i_3}^2W_{i_4}^2W_{i_5}^2W_{i_6}^2W_{j_3}^2W_{j_4}^2W_{j_5}^2W_{j_6}^2\\   \label{c6w2}
&=&\frac{p_0^{11}}{\binom{n}{6}^2}\|W\|_2^{20}O(1),
\end{eqnarray}

\begin{eqnarray}\nonumber
&& \mathbb{E}\Big[\frac{1}{\binom{n}{6}}\sum_{1\leq i_1<\dots<i_6\leq n}\big(A_{i_1i_2}-W_{i_1}W_{i_2}p_0\big)\big(A_{i_2i_3}-W_{i_2}W_{i_3}p_0\big)\\   \nonumber
&&\times \big(A_{i_3i_4}-W_{i_3}W_{i_4}p_0\big)
\big(A_{i_4i_5}-W_{i_4}W_{i_5}p_0\big)\big(A_{i_5i_6}-W_{i_5}W_{i_6}p_0\big)W_{i_6}W_{i_1}p_0 \Big]^2\\    \label{c6w3}
&=&\frac{p_0^7}{\binom{n}{6}^2}\sum_{1\leq i_1<\dots<i_6\leq n}W_{i_1}^3W_{i_6}^3W_{i_2}^2W_{i_3}^2W_{i_4}^2W_{i_5}^2=\frac{p_0^7}{\binom{n}{6}^2}\|W\|_2^{12}O(1)
\end{eqnarray}

\begin{eqnarray}\nonumber
&& \mathbb{E}\Big[\frac{1}{\binom{n}{6}}\sum_{1\leq i_1<\dots<i_6\leq n}\big(A_{i_1i_2}-W_{i_1}W_{i_2}p_0\big)\big(A_{i_2i_3}-W_{i_2}W_{i_3}p_0\big)\\   \nonumber
&&\big(A_{i_3i_4}-W_{i_3}W_{i_4}p_0\big)\\   \label{c6w4}
&&\times\big(A_{i_4i_5}-W_{i_4}W_{i_5}p_0\big)W_{i_5}W_{i_6}p_0W_{i_6}W_{i_1}p_0 \Big]^2\\   \nonumber
&=&\frac{p_0^8}{\binom{n}{6}^2}\sum_{1\leq i_1<\dots<i_6\leq n,i_5<j_6}W_{i_1}^3W_{i_5}^3W_{i_2}^2W_{i_3}^2W_{i_4}^2W_{i_6}^2W_{j_6}^2\\   \nonumber
&=&\frac{p_0^8}{\binom{n}{6}^2}\|W\|_2^{14}O(1)
\end{eqnarray}

\begin{eqnarray}\nonumber
&& \mathbb{E}\Big[\frac{1}{\binom{n}{6}}\sum_{1\leq i_1<\dots<i_6\leq n}\big(A_{i_1i_2}-W_{i_1}W_{i_2}p_0\big)\big(A_{i_2i_3}-W_{i_2}W_{i_3}p_0\big)\\  \nonumber
&&\big(A_{i_3i_4}-W_{i_3}W_{i_4}p_0\big)  
 W_{i_4}W_{i_5}p_0W_{i_5}W_{i_6}p_0W_{i_6}W_{i_1}p_0 \Big]^2\\  \nonumber
&=&\frac{p_0^9}{\binom{n}{6}^2}\sum_{1\leq i_1<\dots<i_6\leq n,i_4<j_5<j_6}W_{i_1}^3W_{i_4}^3W_{i_2}^2W_{i_3}^2W_{i_5}^2W_{i_6}^2W_{j_5}^2W_{j_6}^2\\  \label{c6w5}
&&=\frac{p_0^9}{\binom{n}{6}^2}\|W\|_2^{16}O(1)
\end{eqnarray}

\begin{eqnarray}\nonumber
&& \mathbb{E}\Big[\frac{1}{\binom{n}{6}}\sum_{1\leq i_1<\dots<i_6\leq n}\big(A_{i_1i_2}-W_{i_1}W_{i_2}p_0\big)\big(A_{i_2i_3}-W_{i_2}W_{i_3}p_0\big)\\ \nonumber
&&\times W_{i_3}W_{i_4}p_0 W_{i_4}W_{i_5}p_0W_{i_5}W_{i_6}p_0W_{i_6}W_{i_1}p_0 \Big]^2\\ \nonumber 
&=&\frac{p_0^{10}}{\binom{n}{6}^2}\sum_{1\leq i_1<\dots<i_6\leq n,i_3<j_4<j_5<j_6}W_{i_1}^3W_{i_3}^3W_{i_2}^2W_{i_4}^2\\  \label{c6w6}
&&\times W_{i_5}^2W_{i_6}^2W_{j_4}^2W_{j_5}^2W_{j_6}^2=\frac{p_0^{10}}{\binom{n}{6}^2}\|W\|_2^{18}O(1)
\end{eqnarray}

By (\ref{c3sc61}), (\ref{hatc3s}), (\ref{c3c3}) and  (\ref{c6w1})-(\ref{c6w6}), it yields
\begin{eqnarray}\label{c3sc62}
\widehat{C}_3^2-\widehat{C}_6=2C_3(\widehat{C}_3-C_3)+(C_3^2-C_6)+o_p\big(\sigma_nC_3\big),
\end{eqnarray}
where $\sigma_n^2=\mathbb{E}\big[\widehat{C}_3-C_3\big]^2=\frac{p_0^3}{3!\binom{n}{3}^2}\|W\|_2^6(1+o(1))$. Note that $\sigma_n^2=\frac{C_3}{\binom{n}{3}}(1+o(1))$ by (\ref{c3w}).

By (\ref{c3sc62}), we have
\begin{eqnarray}\nonumber
\frac{\sqrt{\binom{n}{3}}\Big(\widehat{C}_3^2-\widehat{C}_6\Big)}{2C_3\sqrt{C_3}}&=&\frac{\sqrt{\binom{n}{3}}\Big(\widehat{C}_3-C_3\Big)}{\sqrt{C_3}}\\  \label{scac3c6}
&+&\frac{\sqrt{\binom{n}{3}}\Big(C_3^2-C_6\Big)}{2C_3\sqrt{C_3}}+o_p(1).
\end{eqnarray}
Let 
\[\lambda_n=\frac{\sqrt{\binom{n}{3}}\Big(C_3^2-C_6\Big)}{2C_3\sqrt{C_3}}.\]

Under $H_0$, it's easy to check that $\lambda_n=O(np_0^3)=o(1)$. Then it yields by (\ref{hatcc3}), 
\begin{eqnarray}\label{sc3c6}
\frac{\sqrt{\binom{n}{3}}\Big(\widehat{C}_3^2-\widehat{C}_6\Big)}{2C_3\sqrt{C_3}}=\frac{\sqrt{\binom{n}{3}}\Big(\widehat{C}_3-C_3\Big)}{\sqrt{C_3}}+o_p(1)=Y_n+o_p(1),
\end{eqnarray}
where
\begin{eqnarray*}
Y_n=\frac{1}{\sqrt{\binom{n}{3}C_3}}\sum_{1\leq i<j<k\leq n}\big(A_{ij}-W_iW_jp_0\big)\big(A_{jk}-W_jW_kp_0\big)\big(A_{ki}-W_kW_ip_0\big).
\end{eqnarray*}
$Y_n$ converges to the standard normal distribution in law by the following Lemma \ref{lemma1}, which completes the proof of Theorem \ref{tm1}.

\begin{Lemma}\label{lemma1}
Under the condition of Theorem \ref{tm1}, $Y_n$ converges in law to the standard normal distribution if $H_0$ holds.
\end{Lemma}

 The proof relies on the following Martingale central limit theorem. 
\begin{Proposition}[\cite{HH14}]\label{martingale}
 Suppose that for every $n\in\mathbb{N}$ and $k_n\rightarrow\infty$ the random variables $X_{n,1},\dots,X_{n,k_n}$ are a martingale difference sequence relative to an arbitrary filtration $\mathcal{F}_{n,1}\subset\mathcal{F}_{n,2}$ $\subset$ $\dots$ $\mathcal{F}_{n,k_n}$. If (I) $\sum_{i=1}^{k_n}\mathbb{E}(X_{n,i}^2|\mathcal{F}_{n,i-1})\rightarrow 1$ in probability,
 (II) $\sum_{i=1}^{k_n}\mathbb{E}(X_{n,i}^2I[|X_{n,i}|>\epsilon]|\mathcal{F}_{n,i-1})\rightarrow 0$ in probability for every $\epsilon>0$,
\noindent then $\sum_{i=1}^{k_n}X_{n,i}\rightarrow N(0,1)$ in distribution.
 \end{Proposition}

\noindent
\textit{Proof of Lemma \ref{lemma1}}:
Let $X_t=Y_t-Y_{t-1}$ for $3\leq t\leq n$ and $Y_2=0$. Define $F_t=\{A_{ij}|1\leq i<j\leq t\}$. Note that
\begin{eqnarray*}
&&\mathbb{E}(X_t|F_{t-1})=\frac{1}{\sqrt{\binom{n}{3}C_3}}\sum_{1\leq i<j<k=t}\big(A_{ij}-W_iW_jp_0\big)\\
&&\times\mathbb{E}\big[\big(A_{jt}-W_jW_tp_0\big)\big(A_{ti}-W_tW_ip_0\big)\big]=0.
\end{eqnarray*}
Then $X_t$ is a martingale difference.

\noindent
{\bf Check Condition (I)}: It suffices to show that
\begin{equation}\label{exsquare1}
\mathbb{E}\Big[\sum_{t=3}^n\mathbb{E}\big(X_t^2|F_{t-1}\big)\Big]\rightarrow1,\hskip 1cm
\mathbb{E}\Big[\sum_{t=3}^n\mathbb{E}\big(X_t^2|F_{t-1}\big)\Big]^2\rightarrow1.
\end{equation}
Firstly note that
\begin{eqnarray*}
\mathbb{E}\Big[\sum_{t=3}^n\mathbb{E}\big(X_t^2|F_{t-1}\big)\Big]
&=&\sum_{t=3}^n\big(\mathbb{E}Y_t^2-\mathbb{E}Y_{t-1}^2\big)=\mathbb{E}Y_n^2=1.
\end{eqnarray*}

\begin{eqnarray*}
&&\mathbb{E}(X_t^2|F_{t-1})\\
&=&\frac{1}{\binom{n}{3}C_3}\sum_{\substack{1\leq i<j<k=t\\ 1<i_1<j_1<k_1=t}}\mathbb{E}\Big[\big(A_{ij}-W_iW_jp_0\big)\big(A_{jt}-W_jW_tp_0\big)\big(A_{ti}-W_tW_ip_0\big)\\
&&\times \big(A_{i_1j_1}-W_{i_1}W_{j_1}p_0\big)\big(A_{j_1t}-W_{j_1}W_tp_0\big)\big(A_{ti_1}-W_tW_{i_1}p_0\big)|F_{t-1}\Big]\\
&=&\frac{1}{\binom{n}{3}C_3}\sum_{\substack{1\leq i<j<k=t\\ }}\big(A_{ij}-W_iW_jp_0\big)^2W_jW_tp_0\big(1-W_jW_tp_0\big)\\
&&\times W_tW_ip_0\big(1-W_tW_ip_0\big)\\
\end{eqnarray*}

Then
\begin{eqnarray*}\label{exsqu}
&&\mathbb{E}\Big[\sum_{t=3}^n\mathbb{E}\big(X_t^2|F_{t-1}\big)\Big]^2\\
&=&\frac{1}{\binom{n}{3}^2C_3^2}\sum_{\substack{1\leq i<j<t\leq n\\ 1\leq i_1<j_1<t_1\leq n}}\mathbb{E}\Big[\big(A_{ij}-W_iW_jp_0\big)^2W_jW_tp_0\big(1-W_jW_tp_0\big)\\
&&\times W_tW_ip_0\big(1-W_tW_ip_0\big)\big(A_{i_1j_1}-W_{i_1}W_{j_1}p_0\big)^2\\
&&\times W_{j_1}W_{t_1}p_0\big(1-W_{j_1}W_{t_1}p_0\big)W_{t_1}W_{i_1}p_0\big(1-W_{t_1}W_{i_1}p_0\big)\Big]
\end{eqnarray*}

If $\{i,j\}=\{i_1,j_1\}$ in the summation term, then
\begin{eqnarray*}\label{exsqu}
&&\frac{1}{\binom{n}{3}^2C_3^2}\sum_{\substack{1\leq i<j<t\leq n\\ 1\leq i<j<t_1\leq n}}\Big[W_iW_jp_0\big(1-W_iW_jp_0\big)\\
&&\times W_jW_tp_0\big(1-W_jW_tp_0\big)W_tW_ip_0\big(1-W_tW_ip_0\big)\\
&&\times W_{j}W_{t_1}p_0\big(1-W_{j}W_{t_1}p_0\big)W_{t_1}W_{i}p_0\big(1-W_{t_1}W_{i}p_0\big)\Big]\\
&\leq&\frac{p_0^5}{\binom{n}{3}^2C_3^2}\sum_{\substack{1\leq i<j<t\leq n\\ 1\leq i<j<t_1\leq n}}W_{i}^3W_j^3W_t^2W_{t_1}^2\leq \frac{cp_0^5\|W\|_2^8}{\binom{n}{3}^2C_3^2}\\
&& =\frac{1}{p_0\|W\|_2^4}=o(1),
\end{eqnarray*}
if $p_0\|W\|_2^4\rightarrow\infty$.

If $|\{i,j\}\cap\{i_1,j_1\}|=1$ in the summation term, for example, $i=i_1$, then

\begin{eqnarray*}\label{exsqu}
&&\frac{1}{\binom{n}{3}^2C_3^2}\sum_{\substack{1\leq i<j<t\leq n\\ 1\leq i<j_1<t_1\leq n}}\Big[W_iW_jp_0\big(1-W_iW_jp_0\big)\\
&&\times W_jW_tp_0\big(1-W_jW_tp_0\big)W_tW_ip_0\big(1-W_tW_ip_0\big)\\
&&\times W_iW_{j_1}p_0\big(1-W_iW_{j_1}p_0\big)W_{j_1}W_{t_1}p_0\big(1-W_{j_1}W_{t_1}p_0\big)\\
&&\times W_{t_1}W_{i}p_0\big(1-W_{t_1}W_{i}p_0\big)\Big]\\
&\leq&\frac{p_0^6}{\binom{n}{3}^2C_3^2}\sum_{\substack{1\leq i<j<t\leq n\\ 1\leq i<j_1<t_1\leq n}}W_{i}^4W_j^2W_t^2W_{j_1}^2W_{t_1}^2\\
&&\leq \frac{cp_0^6\|W\|_2^{10}}{\binom{n}{3}^2C_3^2}=\frac{1}{\|W\|_2^2}=o(1),
\end{eqnarray*}
if $\|W\|_2\rightarrow\infty$.

If $|\{i,j\}\cap\{i_1,j_1\}|=0$ in the summation term, for example, then
\begin{eqnarray*}\label{exsqu}
&&\frac{1}{\binom{n}{3}^2C_3^2}\sum_{\substack{1\leq i<j<t\leq n\\ 1\leq i_1<j_1<t_1\leq n\\\{i,j\}\cap\{i_1,j_1\}=\o}}\Big[W_iW_jp_0\big(1-W_iW_jp_0\big)\\
&&\times W_jW_tp_0\big(1-W_jW_tp_0\big)W_tW_ip_0\big(1-W_tW_ip_0\big)\\
&&\times W_{i_1}W_{j_1}p_0\big(1-W_{i_1}W_{j_1}p_0\big)W_{j_1}W_{t_1}p_0\big(1-W_{j_1}W_{t_1}p_0\big)\\
&&\times W_{t_1}W_{i_1}p_0\big(1-W_{t_1}W_{i_1}p_0\big)\Big]\\
&=&\frac{1}{\binom{n}{3}^2C_3^2}\sum_{\substack{1\leq i<j<t\leq n\\ 1\leq i_1<j_1<t_1\leq n\\}}\Big[W_iW_jp_0\big(1-W_iW_jp_0\big)W_jW_tp_0\\
&&\times \big(1-W_jW_tp_0\big)W_tW_ip_0\big(1-W_tW_ip_0\big)\\
&&\times W_{i_1}W_{j_1}p_0\big(1-W_{i_1}W_{j_1}p_0\big)W_{j_1}W_{t_1}p_0\big(1-W_{j_1}W_{t_1}p_0\big)\\
&&\times W_{t_1}W_{i_1}p_0\big(1-W_{t_1}W_{i_1}p_0\big)\Big] +\frac{p_0^6\|W\|_2^{10}}{\binom{n}{3}^2C_3^2}\\
&=&\frac{p_0^6}{\binom{n}{3}^2C_3^2}\Big(\sum_{\substack{1\leq i<j<t\leq n\\}}W_{i}^2W_j^2W_t^2\Big)^2+o(1)= 1+o(1),
\end{eqnarray*}
if $\|W\|_2\rightarrow\infty$.
Now (\ref{exsquare1}) holds.

\noindent
{\bf Check Condition (II)}:
For any $\epsilon>0$ and fixed constant $C>0$,
by Cauchy–Schwarz inequality and Markov inequality,  the following inequalities and identitied hold.
\begin{eqnarray}\nonumber
&&\mathbb{E}\Big[\sum_{t=3}^n\mathbb{E}\big(X_t^2I[|X_t|>\epsilon]|F_{t-1}\big)\Big]   \\  \nonumber
&\leq& \mathbb{E}\Big[\sum_{t=3}^n\sqrt{\mathbb{E}\big(X_t^4|F_{t-1}\big)\mathbb{E}\big(I[|X_t|>\epsilon]|F_{t-1}\big)}\Big]\\  \nonumber
&\leq &\frac{1}{\epsilon^2}\mathbb{E}\Big[\sum_{t=3}^n\sqrt{\mathbb{E}\big(X_t^4|F_{t-1}\big)}\sqrt{\mathbb{E}\big(X_t^4|F_{t-1}\big)}\Big]
=\frac{1}{\epsilon^2}\sum_{t=3}^n\mathbb{E}\big(X_{t}^4\big)\\     \nonumber
&=&\frac{1}{\epsilon^2\binom{n}{3}^2C_3^2}\sum_{\substack{1\leq i<j<t\leq n \\1\leq i_1<j_1<t\leq n \\1\leq i_2<j_2<t\leq n \\1\leq i_3<j_3<t\leq n }}\mathbb{E}\Big[\big(A_{ij}-W_iW_jp_0\big)\big[\big(A_{jt}-W_jW_tp_0\big)\\   \nonumber
&&\times \big(A_{ti}-W_tW_ip_0\big)\\   \nonumber
&&\times \big(A_{i_1j_1}-W_{i_1}W_{j_1}p_0\big)\big(A_{j_1t}-W_{j_1}W_tp_0\big)\big(A_{ti_1}-W_tW_{i_1}p_0\big)\\    \nonumber
&&\times \big(A_{i_2j_2}-W_{i_2}W_{j_2}p_0\big)\big(A_{j_2t}-W_{j_2}W_tp_0\big)\big(A_{ti_2}-W_tW_{i_2}p_0\big)\\    \nonumber
&&\times \big(A_{i_3j_3}-W_{i_3}W_{j_3}p_0\big)\big[\big(A_{j_3t}-W_{j_3}W_tp_0\big)\big(A_{ti_3}-W_tW_{i_3}p_0\big)\Big]\\    \nonumber
\end{eqnarray}
\begin{eqnarray}
&=&\frac{C}{\epsilon_1^2\binom{n}{3}^2C_3^2}\sum_{\substack{1\leq i<j<t\leq n\\1\leq i_1<j_1<t\leq n}}\mathbb{E}\Big[\big(A_{ij}-W_iW_jp_0\big)^2\big[\big(A_{jt}-W_jW_tp_0\big)^2\\ \nonumber
&&\times \big(A_{ti}-W_tW_ip_0\big)^2\\   \nonumber
&&\times \big(A_{i_1j_1}-W_{i_1}W_{j_1}p_0\big)^2\big(A_{j_1t}-W_{j_1}W_tp_0\big)^2\big(A_{ti_1}-W_tW_{i_1}p_0\big)^2\Big]\\   \nonumber
&&+\frac{C}{\epsilon^2\binom{n}{3}^2C_3^2}\sum_{\substack{1\leq i<j<t\leq n}}\mathbb{E}\Big[\big(A_{ij}-W_iW_jp_0\big)^4\big(A_{jt}-W_jW_tp_0\big)^4\\ \nonumber
&&\times \big(A_{ti}-W_tW_ip_0\big)^4\Big]\\   \nonumber
&=&\frac{Cp_0^6\|W\|_2^{10}}{\epsilon^2\binom{n}{3}^2C_3^2}+\frac{Cp_0^5\|W\|_2^{8}}{\epsilon^2\binom{n}{3}^2C_3^2}+\frac{Cp_0^3\|W\|_2^{6}}{\epsilon^2\binom{n}{3}^2C_3^2}\\   \nonumber
&=&O\Big(\frac{1}{\|W\|_2^2}+\frac{1}{p_0\|W\|_2^4}+\frac{1}{p_0^3\|W\|_2^6}\Big)\rightarrow0,
\end{eqnarray}
if $p_0\|W\|_2^2\rightarrow\infty$, $\|W\|_2\rightarrow\infty$.
Hence, $\sum_{t=m}^n\mathbb{E}\big(Z_t^2I[|Z_t|>\epsilon]|F_{t-1}\big)=o_P(1)$. By Proposition \ref{martingale}, the proof is complete.

\subsection{Proof of Theorem \ref{tm2}} 
Under $H_1$, one has
\[\lambda_n^2\asymp\frac{(a-b)^6b^6r^6(1-r)^6\|W\|_2^6}{n^{12}p_0^9}.\]
If $a\asymp b$, then
\[\lambda_n^2\asymp \frac{r^6(a-b)^6\|W\|_2^6}{n^3b^3}.\]

By a similar proof of Theorem \ref{tm1}, equation (\ref{scac3c6}) still holds under $H_1$. The first term and the third terms in (\ref{scac3c6}) are bounded in probability. Under condition (\ref{powerc}), the middle term in equation (\ref{scac3c6}) tends to infinity. Then the proof is complete. 
\qed


\begin{thebibliography}{99}

\bibitem{ACL01}
Aiello, W., Chung, F. and Lu, L. (2001). 
A random graph model for power law graphs.
\textit{Experimental Mathematics}, 20, 53-66.

%\bibitem{A18} Abbe, E. (2018).
%Community detection and stochastic block models: recent developments. 
%\emph{Journal of Machine Learning Research}, 18, 6446-6531. 


%\bibitem{ACB13} Amini, A., Chen, A. and Bickel, P. (2013).
%Pseudo-likelihood methods for community detection in large sparse networks.
%\textit{Annals of Statistics}, \textbf{41(4)}, 2097-2122.

%\bibitem{BS16} Bickel, P. J. and Sarkar, P. (2016). Hypothesis testing for automated community detection in networks.
%\textit{Journal of Royal Statistical Society, Series B}, \textbf{78}, 253-273.



\bibitem{AC09}
Anderson, R. and Chellapilla, K. 2009.
Finding dense subgraphs with size bounds.
\textit{WAW 2009}, 25-37.





\bibitem{AV14}
Arias-Castro, E. and N. Verzelen. 2014.
Community detection in dense random networks.
\textit{Annals of Statistics}, 42, 3: 940-969.





\bibitem{BCHV19}
Bogerd, K. et al. 2019.
Cummunity detection in inhomogeneous random graphs.
\url{https://arxiv.org/abs/1909.03217}




\bibitem{BJR07}
Bollobas, B., Janson, S. and Riordan, O. 2007.
The phase transition in inhomogeneous random graphs.
\textit{Random Structures \& Algorithms}, 31(1): 3-122.

\bibitem{C00}
 Charikar, M. 2000. Greedy approximation algorithms for finding dense components in a graph. 
 \textit{APPROX}, 84–95.


%\bibitem{CK10}Chertok, M. and Y. Keller. 2010.
%Efficient high order matching. 
%\textit{IEEE Trans. on Pattern Analysis and
%Machine Intelligence}, \textbf{32(12)}, 2205-2215.

\bibitem{CY06}Chen, J. and  B. Yuan. 2006.
Detecting functional modules in the yeast proteinprotein interaction
network. \textit{Bioinformatics}, \textbf{22(18)}, 2283-2290.


\bibitem{CS10}
Chen, J. and Y. Saad. 2010.
Dense subgraph extraction with application to community detection,
\textit{IEEE Transactions on Knowledge and Data Engineering} 24,7: 1216-1230.

%\bibitem{DHM17}
%Dolezal, M., Hladky, J. and Mathe, A.(2017).
%Cliques in dense inhomogeneous random graphs.
%\textit{Random Structures \& Algorithms}, 51(2): 275-314.


\bibitem{ER60}Erd\"{o}s, P. and A. R\'{e}nyi. 1960.
 On the evolution of random graphs. \textit{Publ. Math. Inst. Hungar. Acad.
Sci. }, \textbf{5}, 17-61.


\bibitem{GKT05}
Gibson, D., R. Kumar, and A. Tomkins. 2005. Discovering large dense subgraphs in massive graphs. 
\textit{VLDB} 05,721–732.

%\bibitem{GL17a} Gao, C. and J. Lafferty. 2017.
%Testing for global network structure using small subgraph statistics,
%\url{https://arxiv.org/pdf/1710.00862.pdf}



%\bibitem{DKMZ11} Decelle, A., F. Krzakala, C. Moore, and F. Zdeborov\'{a}. 2011. Asymptotic
%analysis of the stochastic block model for modular networks and its
%algorithmic applications. \textit{Physics Review E}, \textbf{84}, 066-106.




%\bibitem{F10} Fortunato, S. 2010. Community detection in graphs. \textit{Physics Reports}, \textbf{486 (3-5)},
%75-174.




\bibitem{GKT05}
Gibson, D., R. Kumar, and A. Tomkins. 2005.
Discovering large dense subgraphs in massive
graphs.
\textit{VLDB 05},721–732.



%\bibitem{GL17a} Gao, C. and J. Lafferty. 2017.
%Testing Network Structure Using Relations Between
%Small Subgraph Probabilities,
%\url{https://arxiv.org/pdf/1704.06742.pdf}



%\bibitem{GZFA10} Goldenberg, A., A. X. S.  Zheng, E. Fienberg, and E. M., Airoldi. 2010.
%A survey of statistical network models. 
%\textit{Foundations and Trends in Machine Learning 2}, \textbf{2}, 129-233.
\bibitem{HWX2018}
Hajek, B., Wu, Y. and Xu, J. 2018.
Recovering a hidden community beyond the Kesten Stigum threshold in $O(|E|\log^{\*}|V|)$ time. 
\textit{Journal of Applied Probability}, 55, 2: 325-352.

\bibitem{HH14} Hall, P. and C. C., Heyde. 2014.
\textit{Martingale limit theory and its application}. Academic press.






\bibitem{HSBSSF16}
Hooi,B., H.A., Song, A. Beutel, N. Shah, K. Shin,and C. Faloutsos. 2016.
Fraudar:bounding graph fraud in the face of camouflage,
\textit{KDD,ACM},895-904.




%\bibitem{HWX18}
%Hajek, B., Y. Wu, and J. Xu. 2018.
%Recovering a hidden community beyond the Kesten-Stigum threshold in $O(|E|\log^{\*}|V|)$ time. \textit{Journal of Applied Probability}, 55, 2: 325-352.




%\bibitem{JKL18}
%Jin, J., Ke, T. and Luo, S. 2018. Network global testing by counting graphlets.
%\textit{Proceedings of the 35th International Conference on Machine Learning}, 80:2333-2341.



\bibitem{KS09}
Khuller, S. and Saha, B. 2009.
On finding dense subgraphs,
\textit{ICALP 2009}: 597-608. 









\bibitem{VA15}
Verzelen, N., and E. Arias-Castro. 2015.
Community detection in sparse random networks. 
\textit{Ann. Appl. Probab.} 25,6:3465--3510. 



\bibitem{WLKN09}
Wu,M., X. Li, C.K. Kwoh, and S.K. Ng. 2009.
A coreattachment
based method to detect protein complexes in
ppi networks, 
\textit{BMC bioinformatics}, 10: 169.










\end{thebibliography}
\end{document}